\documentclass[showpacs,twocolumn,superscriptaddress]{revtex4-2}
\usepackage[utf8]{inputenc}
\usepackage{hyperref}
\hypersetup{colorlinks=true,linkcolor=blue,citecolor=cyan}
\usepackage{array,multirow}
\def\be{\begin{equation}}
	\def\ee{\end{equation}}
\def\bestar{\begin{equation*}}
	\def\eestar{\end{equation*}}

\usepackage{graphicx}
\usepackage{dcolumn}
\usepackage{bm}
\usepackage{color}
\usepackage{enumitem}
\usepackage{amsmath}
\usepackage{amssymb}
\usepackage{adjustbox}
\usepackage{graphicx}
\usepackage{soul,xcolor}
\setstcolor{red}

\begin{document}
\title{Radiation properties of the accretion disk around a black hole in Einstein-Maxwell-scalar theory}

\author{Mirzabek Alloqulov}
\email{malloqulov@gmail.com}
\affiliation{New Uzbekistan University, Movarounnahr street 1, Tashkent 100000, Uzbekistan}
\affiliation{Institute of Fundamental and Applied Research, National Research University TIIAME, Kori Niyoziy 39, Tashkent 100000, Uzbekistan} 
\affiliation{University of Tashkent for Applied Sciences, Str. Gavhar 1, Tashkent 100149, Uzbekistan}

\author{Sanjar Shaymatov}
\email{sanjar@astrin.uz}
\affiliation{Institute for Theoretical Physics and Cosmology, Zhejiang University of Technology, Hangzhou 310023, China}
\affiliation{Central Asian University, Milliy Bog Street 264, Tashkent 111221, Uzbekistan}
\affiliation{Institute of Fundamental and Applied Research, National Research University TIIAME, Kori Niyoziy 39, Tashkent 100000, Uzbekistan}
\affiliation{National University of Uzbekistan, Tashkent 100174, Uzbekistan}

\author{Bobomurat Ahmedov}
\email{ahmedov@astrin.uz}
\affiliation{Institute of Fundamental and Applied Research, National Research University TIIAME, Kori Niyoziy 39, Tashkent 100000, Uzbekistan}
\affiliation{Ulugh Beg Astronomical Institute, Astronomy St 33, Tashkent 100052, Uzbekistan}
\affiliation{Institute of Theoretical Physics, National University of Uzbekistan, Tashkent 100174, Uzbekistan}

\author{Abdul Jawad}
\email{jawadab181@yahoo.com}
\affiliation{Institute for Theoretical Physics and Cosmology, Zhejiang University of Technology, Hangzhou 310023, China}
\affiliation{Department of Mathematics, COMSATS University Islamabad, Lahore Campus, Lahore-54000, Pakistan}

%

\date{\today}
\begin{abstract}
In this study, we explore the properties of a non-rotating black hole in the Einstein-Maxwell-scalar (EMS) theory and investigate the luminosity of the accretion disk surrounding it. We determine all the orbital parameters of particles in the accretion disk, including the radius of the innermost stable circular orbit (ISCO) with angular velocity, angular momentum, and energy. Further, we study the radiative efficiency for different values of black hole parameters. Finally, we analyze the flux, differential luminosity, and temperature of the accretion disk.
\end{abstract}

\maketitle
\footnotesize
\section{Introduction}

Black holes are assumed to form as a result of the
gravitational collapse of compact massive objects that have exhausted all their resources to resist gravity. Therefore, they are fascinating objects with a very attractive nature. The existence of black holes has been confirmed by recent gravitational wave \cite{Abbott16a,Abbott16b} and astronomical observations associated with the BlackHoleCam and EHT collaborations \cite{Akiyama19L1,Akiyama19L6}. These observational studies open up captivating avenues for forthcoming investigations to test remarkable aspects and gravitational properties of black holes in general relativity (GR) and modified theories of gravity.

It is a fact that astrophysical black holes can have three parameters in most cases, including black hole mass $M$, spin $a$, and charge $Q$. According to recent astronomical observations, black holes can be considered as rotating \cite{Bambi17-BHs,Walton13,Patrick11b-Seyfert,Patrick11a-Seyfert,Tan12,Gallo11} as well as charged black holes that correspond to a Reissner-Nordstr\"{o}m (RN) black hole solution characterized by $M$ and $Q$. This well-known RN solution in GR was first proposed by H. Reissner in 1916 \cite{Reissner16}. It was also independently considered by G. Nordstr\"{o}m in 1918 \cite{Nordstrom18} with linear Maxwell electrodynamics.Afterwards, the remarkable aspects of the RN black hole were discussed by addressing the motion of neutral and charged particles~\cite{Pugliese11,Pugliese11b}. Several astrophysical mechanisms have since been proposed in relation to how black holes can possess electric charge. For example, the balance between gravitational and Coulomb forces for charged particles allows black holes to have a positive net electric charge at the compact object's near surface~\cite{Zajacek19,Bally78}, and the irradiating photons also give rise to charged matter at the surface~\cite{Weingartner06}. Another way is that the twisted magnetic field lines due to the black hole spin can produce the induced charge, referring to the well-known Wald mechanism \cite{Wald74}. Later, an exact rotating Schwinger dyon solution was proposed as a charged black hole solution with mass $M$, electric charge $Q_e$, magnetic charge $Q_m$ and spin $a$~\cite{Kasuya82,Shaymatov22b}. Charged black hole solutions that are free of the physical singularity $r = 0$ also exist, which are referred to as regular black hole solutions coupled to non-linear electrodynamics (NED) in GR \cite{Bardeen68,Ayon-Beato98,Ayon-Beato99,Bronnikov01,Bambi13,Fan2016,Balart_2023,rincon2023,Balart_2023b,Kruglov_2023,Kruglov_2022,Kruglov_2021,Panotopoulos_2020,Rincon2021,panotopoulos2019,Rincon2018,Panotopoulos2018} addressing various
situations. As these solutions can avoid physical singularity, they have since been studied and tested for various possible cases~\cite{Stuchlik15,Schee15,Bambi13d}. 

Einstein gravity is known as a best theory in the strong field regime. The GR can also be extended to the low energy limit of the string theory associated with the dilaton scalar field as an extra term added to the action that refers to the Einstein-Hilbert action, including the axion, gauge, and dilaton fields \cite{Green87book,Gibbons88}. In this context, the authors of Ref.~\cite{Garfinkle91} considered the heterotic string theory including the scalar dilaton field that is coupled to the electromagnetic field tensor. On these lines, there has been an extensive analysis~\cite{Koikawa87,Gibbons88,Garfinkle91,Brill91,Harms92,Rakhmanov94} addressing the causal structures and thermodynamic properties of black hole solutions associated with the dilaton field. Several investigations have also been conducted on black hole solutions in various extended theories~\cite{Gubser98,Witten98,Maldacena99,Aharony00,Klemm01}. In the context of extended theories of gravity, some approaches including quantum features have been proposed and analyzed so as to give a peculiarity to the literature (see, for example, in~\cite{Koch_2014,Rincon_2017,Koch_2011,Koch_2016,Bonanno_1999,Bonanno_2021,bonanno2022,Bonanno_2023}). 

From the theoretical and observational studies, the existence of an accretion disk around black holes is expected to be the primary source of information about gravity and geometry in the strong field regime. It can also enable very potent tests for probing unknown aspects of high energy phenomena occurring in the close vicinity of black holes~\cite{Abramowicz13}. Therefore, it has value to consider important and valuable insights of the spacetime geometry in the strong field regime that can have a significant impact on the test particle geodesics and can result in changing observable quantities, that is, the innermost stable circular orbit (ISCO) and accretion disk parameters. With this in view, having confidence with the conclusions from observations related to the accretion disks is significantly important. Thus, observational data on the thin accretion disk though the anticipated thermal spectra could play a decisive role in testing the gravity in the strong field regime. With this motivation in mind, in this work, we explore the properties of an interesting solution describing a non-rotating black hole in the Einstein-Maxwell-scalar (EMS) theory. We investigate remarkable aspects and radiation properties of the accretion disk around a black hole on the basis of the geometrically thin Novikov–Thorne model as an optically thick disk. We also analyze the ISCOs around black holes in the EMS theory. Further, we represent the radiation properties of the accretion disk, that is, the temperature and spectral energy distributions for a thin Novikov–Thorne disk as a consequence of the comparison with the results of the standard Schwarzschild case in the context of GR.  

The paper is organized as follows. We briefly review the solution of the charged black hole in the EMS theory in Sec.~\ref{section2}. Sec.~\ref {section3} is devoted to the study of charged test particle dynamics around the black hole and the geometrically thin Novikov-Thorne model for the accretion disk. We discuss the flux of the radiant energy over the accretion disk, accretions disk's radiative efficiency, temperature profile, and differential luminosity in Sec.~\ref{section4}. We summarize our results in Sec.~\ref{summary}. Throughout this paper, we use the $(–, +, +, +)$ signature for the spacetime metric and system of units in which we set $G=c=1$.

\section{Static black holes in Einstein-Maxwell-scalar theory \label{section2}}

Here we review the charged black hole solutions in EMS theory. To this end, we first write the action as follows~\cite{Gibbons88,Yu_2021}
\begin{eqnarray}\nonumber\label{action}
S=\int d^4x\sqrt{-g}\Big[R-2\nabla_\alpha\phi \nabla^\alpha\phi-K(\phi) F_{\alpha\beta}F^{\alpha\beta}
-V(\phi)
\Big]\, ,
\end{eqnarray}
with the covariant derivative $\nabla_\alpha$, Ricci scalar $R$ and massless scalar field $\phi$. $F_{\alpha\beta}$ and $K(\phi)$ refer to the electromagnetic field tensor and the function of scalar field respectively, that describes the coupling function between the electromagnetic and dilaton fields. $V(\phi)$ refers to the scalar potential associated with the cosmological constant $\Lambda$ in the frame of the EMS theory and describes the black hole in the de-Sitter universe coupled with the dilaton field, that is, $V(\phi)=\Lambda\left(e^{2\phi}+4+e^{-2\phi}\right)/3$ \cite{Gao04}. However, we further restrict ourselves to the black hole solution with vanishing cosmological constant, i.e., $V(\phi)=0$. With this in view, the metric describing a static and spherically symmetric black hole solution in the Schwarzschild coordinate system is then given by~\cite{Yu_2021}
\begin{equation}\label{metric}
    ds^2=-U(r)dt^2+\frac{dr^2}{U(r)}+f(r)\left(d\theta^2+\sin^2\theta d\varphi^2\right)\ ,
\end{equation}
where $U(r)$ and $f(r)$ refer to radial functions given in terms of the function $K(\phi)$ that is written as    
\begin{eqnarray}\label{KV}
K(\phi)=\frac{2e^{2\phi}}{\beta e^{4\phi}+\beta-2\gamma}\, .
\end{eqnarray}
Hence, these two functions are written in the following forms:
 \begin{eqnarray}\label{metfuncts}\nonumber
&& f(r)=r^2\left(1+\frac{\gamma Q^2}{Mr}\right), \\ && U(r)=1-\frac{2M}{r}+\frac{\beta Q^2}{f(r)}\, ,
 \end{eqnarray}
with mass $M$ and electric charge $Q$ of the black hole. We note that the new terms are given with $\beta$ and $\gamma$ which are dimensionless constants of the static black hole solution in EMS theory. The metric functions $f(r)$ and $U(r)$ reduce to the Schwarzschild case in the limit of $\beta=0$ and $\gamma=0$, and to the Reissner-Nordstr\"{o}m case in the limit of $\gamma=0$ and $\beta=1$. However, it is worth noting that the functions $f(r)$ and $U(r)$ present the dilation solution in the limit of $\beta=0$ and $\gamma=-1$ (i.e., $K(\phi)=e^{2\phi}$ and $V(\phi)=0$ \cite{Gibbons88,Garfinkle91}).

We assume that the vector and the dilaton fields depend on the radial coordinate only as shown in Ref~\cite{Yu_2021}, and thus their radial form can be defined by  
\begin{eqnarray}
 && A_t(r)=\frac{Q}{r}\left[\gamma-\frac{\beta}{2}\left(1+\frac{r^2}{f(r)}\right)\right] , 
 \\
 && \phi(r)=-\frac{1}{2}\ln\left(\frac{f(r)}{r^2}\right)\ .
\end{eqnarray}
Further the condition $U(r)=0$ with respect to $r$ solves to give the black hole horizon as 
\begin{equation}\label{horizon}
   \frac{r_h}{M}= 1-\frac{\gamma  Q^2}{2 M^2}+\sqrt{1+\frac{Q^2 (\gamma -\beta )}{M^2}+\frac{\gamma ^2 Q^4}{4 M^4}}\ .
\end{equation}

In Fig.~\ref{metric}, we demonstrate the radial profiles of the radial function $U(r)$ for various combinations of the EMS parameters $\beta$ and $\gamma$ and also for the black hole charge. As can be observed from Fig.~\ref{metric}, the zeros of the metric function imply the location of black hole horizons. Fig.~\ref{metric} shows that Cauchy and event horizons become close to each other for large values of $\beta$. In addition, the black hole event horizon increases for negative values of $\gamma$, whereas it decreases for $\gamma>0$. 

Let us now examine the characteristics of the extremes in the black hole charge and the minimum in the event horizon radius by setting the following condition:
\begin{equation}
    U(r)=U'(r)=0\, .
\end{equation}
The above condition allows one to obtain the radius and charge expressions as follows:
\begin{eqnarray}
\label{rhmin}
  \frac{(r_h)_{min}}{M}=2-\frac{\beta }{\gamma }+ \frac{\sqrt{\beta ^2-2 \beta  \gamma }}{\gamma }\ ,\\ \label{qextr}
\frac{Q_{extr}^2}{M^2} =\frac{2 \left(\beta- \sqrt{\beta^2 -2 \beta \gamma }-\gamma \right)}{\gamma ^2}\, .
\end{eqnarray}

\begin{figure*}[ht]
\centering
\includegraphics[width=0.4\textwidth]{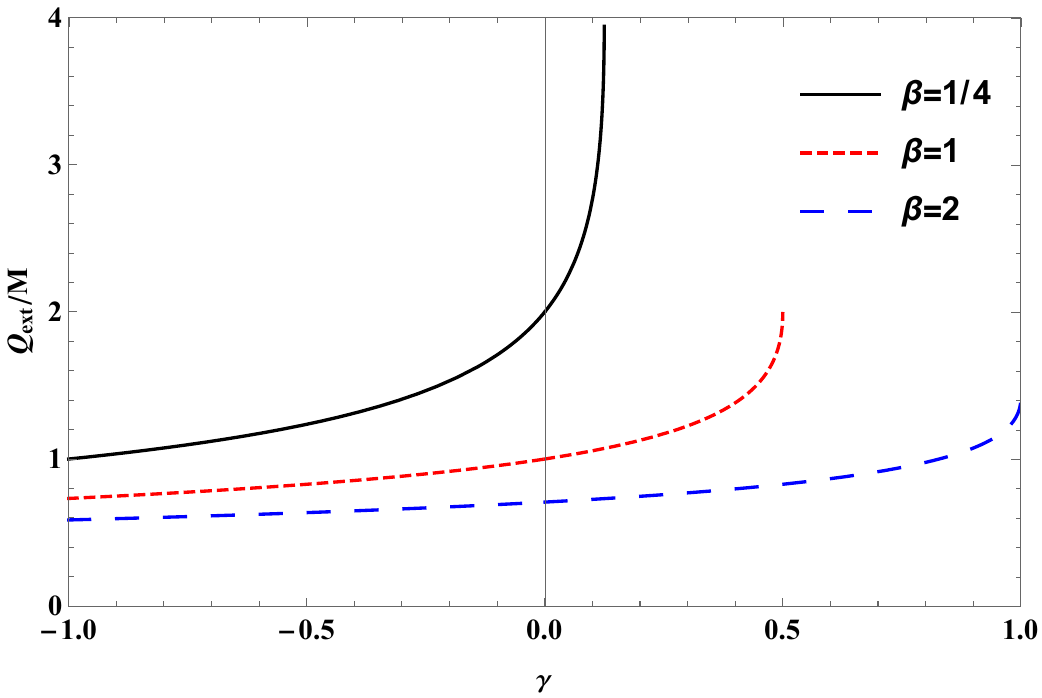}
\includegraphics[width=0.4\textwidth]{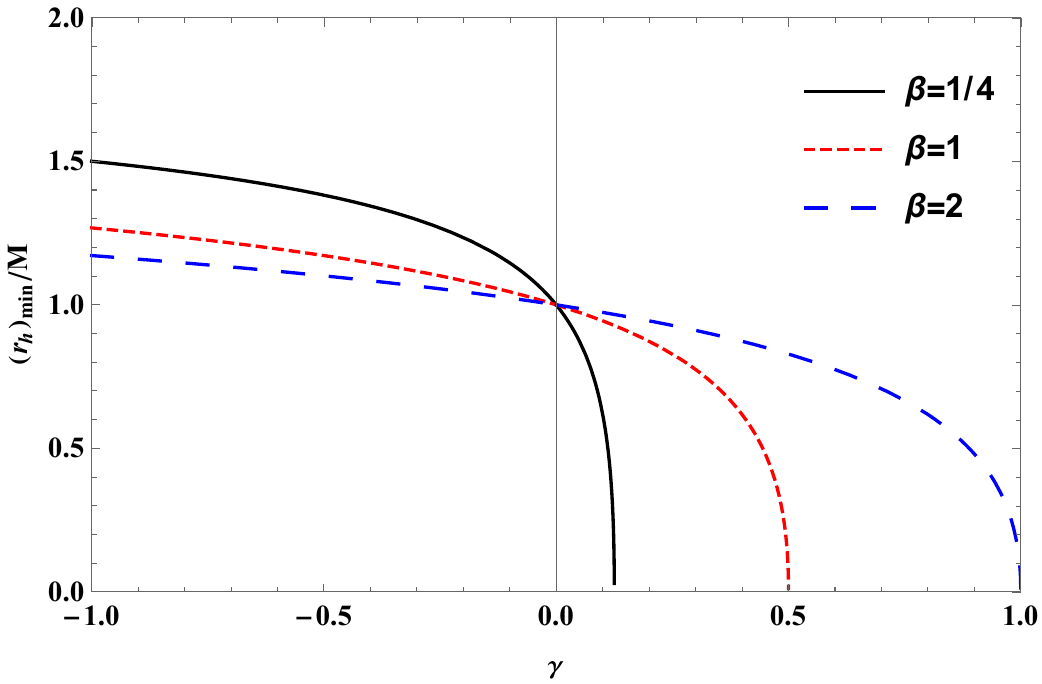}
 \caption{Critical values of black hole charge (left panel) and the minimal radius of the horizon (right panel) as functions of the parameter $\gamma$ for various combinations of $\beta$.} 
 \label{extrem}
 \end{figure*}

In Fig.~\ref{extrem} we show the dependence of the minimum values of the outer horizon and extreme values of the black hole charge from the parameter $\gamma$ for the different values of  $\beta$. The solid lines stand for the minimum event horizon, whereas the dashed ones stand for the extreme charge. As can be observed from Fig.~\ref{extrem}, the minimum radius of the outer horizon decreases with an increase in the value of $\gamma$, and it also decreases with an increase in the value of $\beta$ at $\gamma<0$. Similarly, the extreme in $Q$ also decreases at $\gamma<0$. 
The event horizon becomes zero in the case in which $\beta$ reaches its critical value, thus implying that $\beta$ must be less than $\beta_{cr}$. The critical value of $\beta$ increases with the increase parameter $\gamma$. 

Our numerical analysis shows that for all possible values of parameter $\gamma$, the following conditions are always satisfied: 
$$\lim_{\beta \to 0} \left\{(r_h)_{\rm min}, Q_{\rm extr}\right\}=\{2M,0\}, $$ and $$\lim_{\beta \to \beta_{\rm max}} \left\{(r_h)_{\rm min}, Q_{\rm extr}\right\}=\{0, Q_{\rm extr}\}.$$ 
The upper value of $\beta_{\rm max}$ ($Q_{\rm extr}$) increases (decreases) as parameter $\gamma$ increase. Together with Eqs.~(\ref{rhmin}) and (\ref{qextr}) one can easily find the following relations: 
\begin{equation}
       \beta_{\rm max}=2 \gamma, \qquad \frac{Q_{\rm extr}}{M}=\sqrt{\frac{2}{\gamma}}\ .
   \end{equation}
In Table \ref{Tab1}, we present the critical values of $\beta$ and corresponding values of $Q_{extr}$ for the given values of $\gamma$. 

\begin{table}[ht]
\centering
\begin{tabular}{|c |c | c | c | c | c | c | c |}
\hline
$\gamma$ & $\beta_{\rm max}$ & $Q_{\rm extr}/M$ \\ [1.0ex]
\hline
1/8 & $1/4$ & $4$ \\[1.0ex]
\hline
$1/4$ & $1/2$ & $2\sqrt{2}$ \\[1.0ex]
\hline
$3/8$  & $3/4$ & $4/3$ \\[1.0ex]
\hline
$1/2$  & $1$ & $2$ \\[1.0ex]
\hline
$3/4$  & $3/2$ & $2\sqrt{2}/3$ \\[1.0ex]
\hline
$1$  & $2$ & $\sqrt{2}$ \\[1.0ex]
\hline
\end{tabular}
\caption{Extremes of black hole charge and critical value of $\gamma$ for different values of $\beta$ when the event horizon is zero.}\label{Tab1}
\end{table}
\begin{figure*}[ht]
\centering
\includegraphics[width=0.4\textwidth]{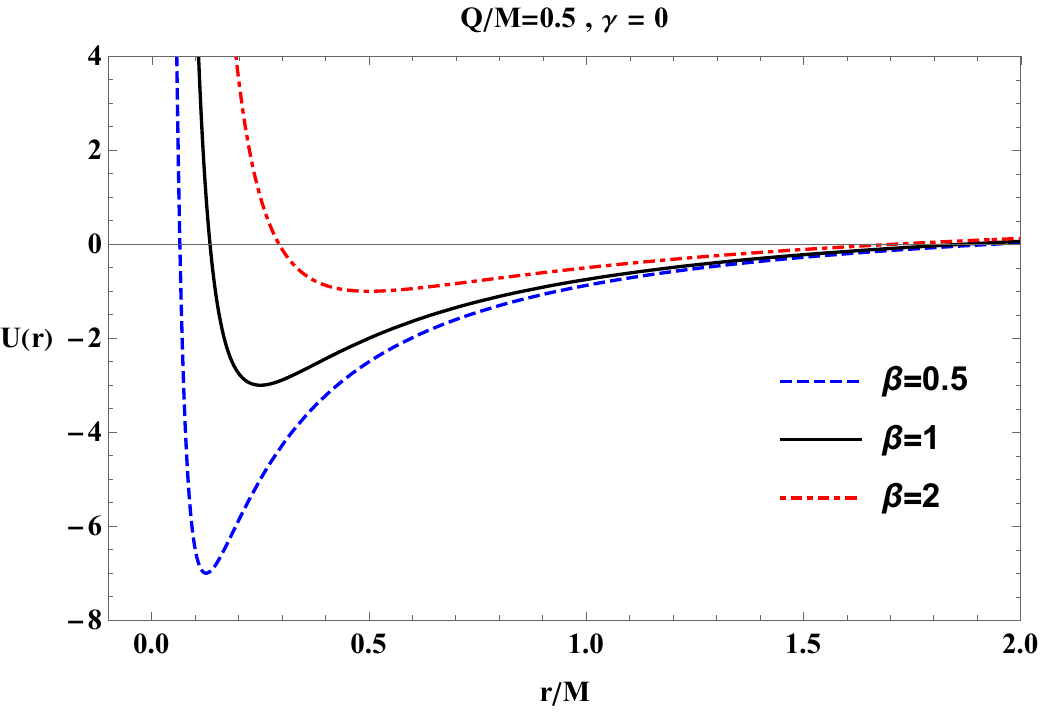}
\includegraphics[width=0.4\textwidth]{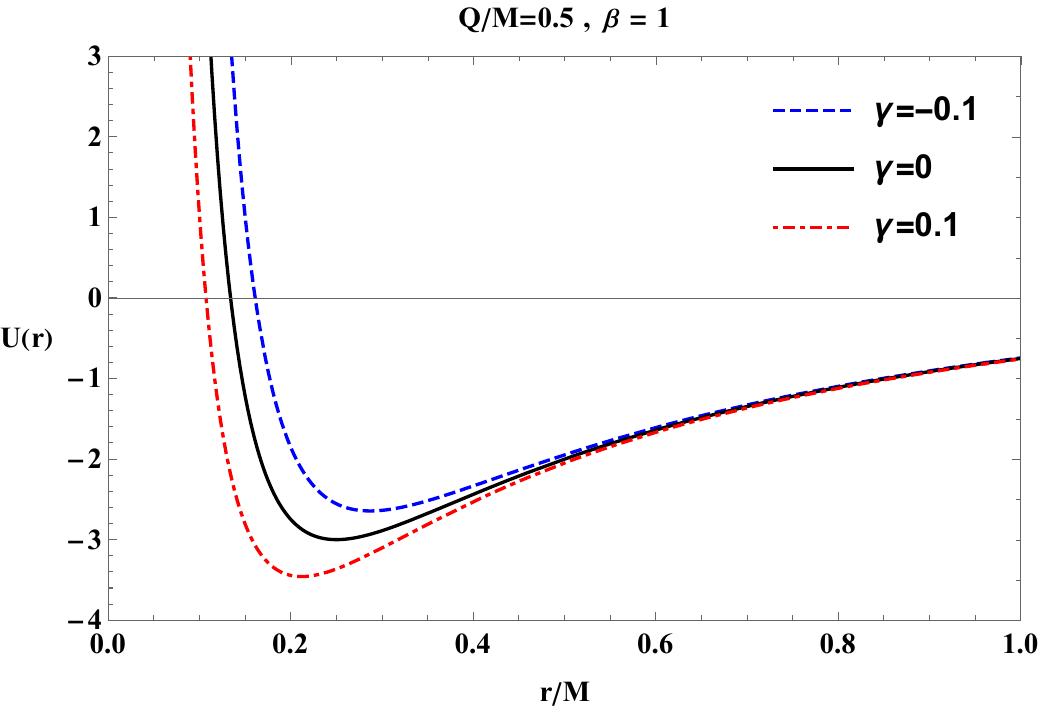}
 \caption{Radial dependence of the metric function $U(r)$ plotted for the different values of parameters $\beta$ and $\gamma$ in the left and right panels, respectively.}  
\label{fig:metric}
 \end{figure*}
\begin{figure*}[ht]
  \includegraphics[width=0.4\textwidth]{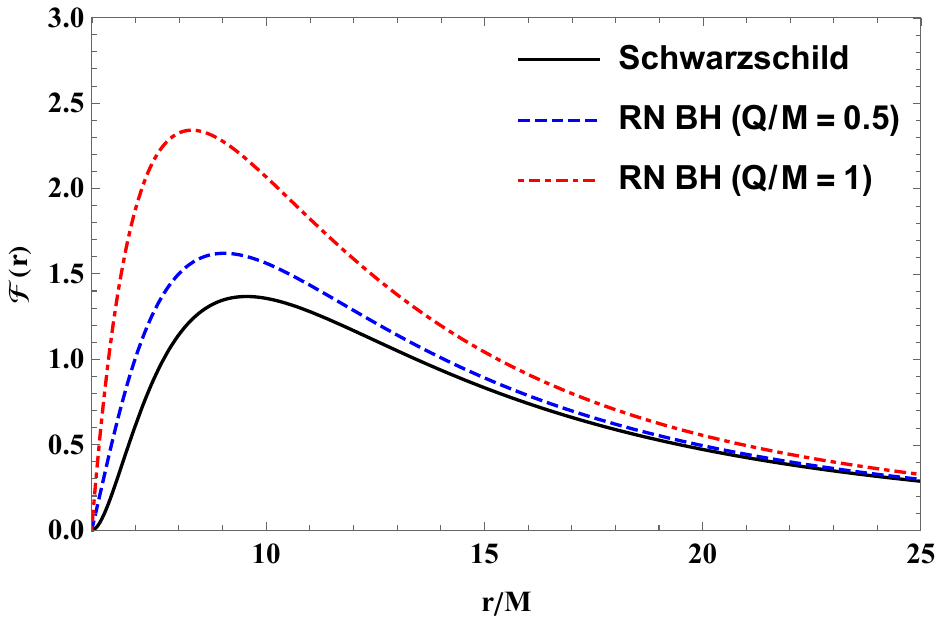}  
  \includegraphics[width=0.4\textwidth]{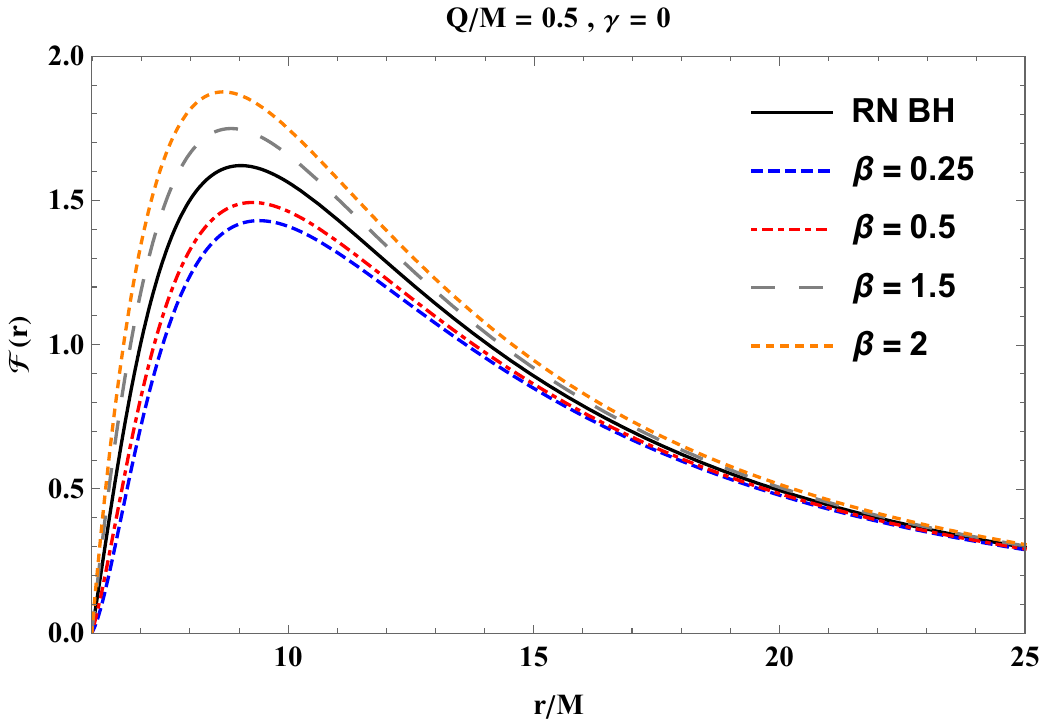}
  \begin{center}
 \includegraphics[scale=0.4]{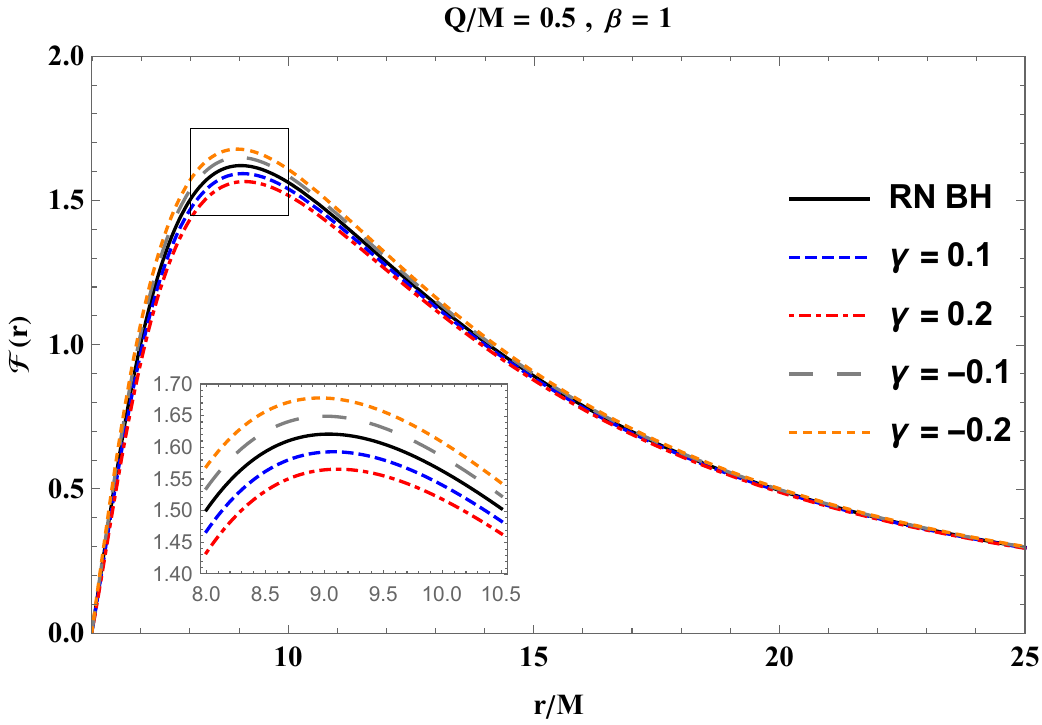}
 \end{center}
\caption{\label{fig:flux} Radial dependence of the flux of electromagnetic radiation of the accretion disk on different values of $Q$, $\beta$, and $\gamma$ plotted for the black hole in EMS theory. Numerical evaluation of the flux $\mathcal{F}$ divided by $10^{-5}$ of the accretion disk as a function of $r/M$}
\end{figure*}  
\begin{figure*}
    \includegraphics[width=0.4\textwidth]{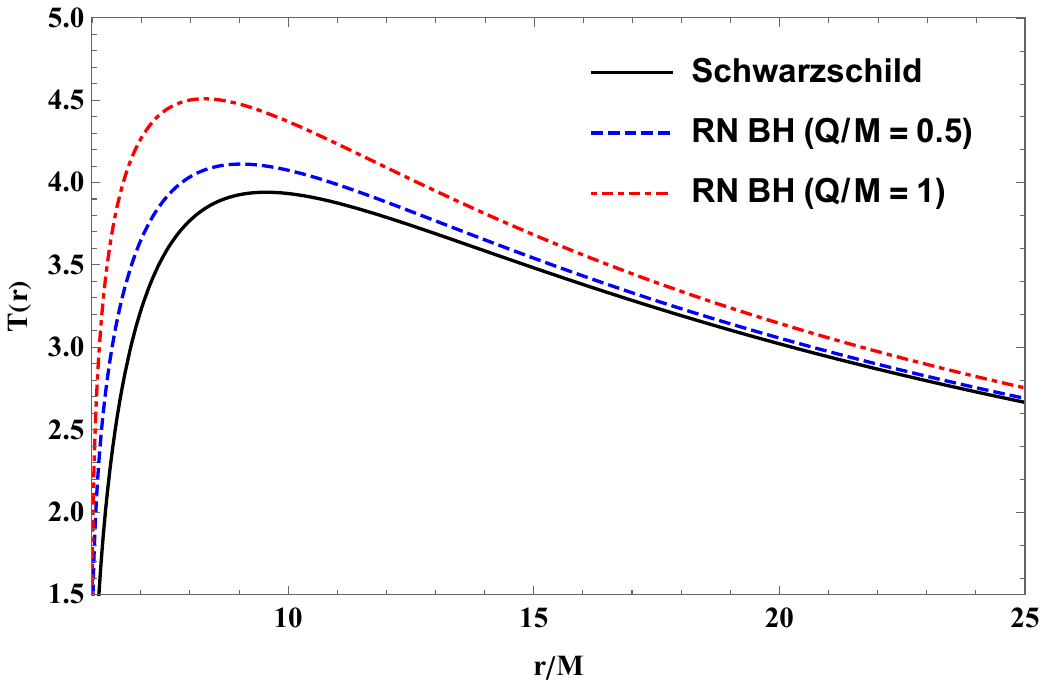}
    \includegraphics[width=0.45\textwidth]{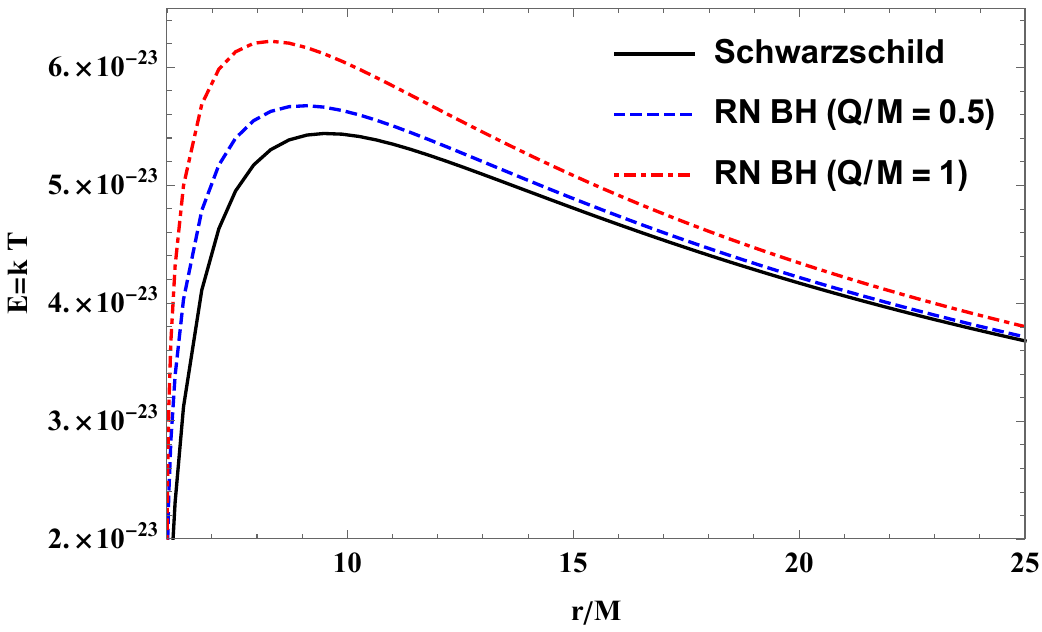}
        \caption{Radial dependence of the temperature and energy of the accretion disk plotted in the left and right panels, respectively,  for different values of black hole charge $Q$ in the case of fixed $\gamma=0$ and $\beta=1$. }
    \label{temperature1}
\end{figure*}
\begin{figure*}
    \includegraphics[width=0.4\textwidth]{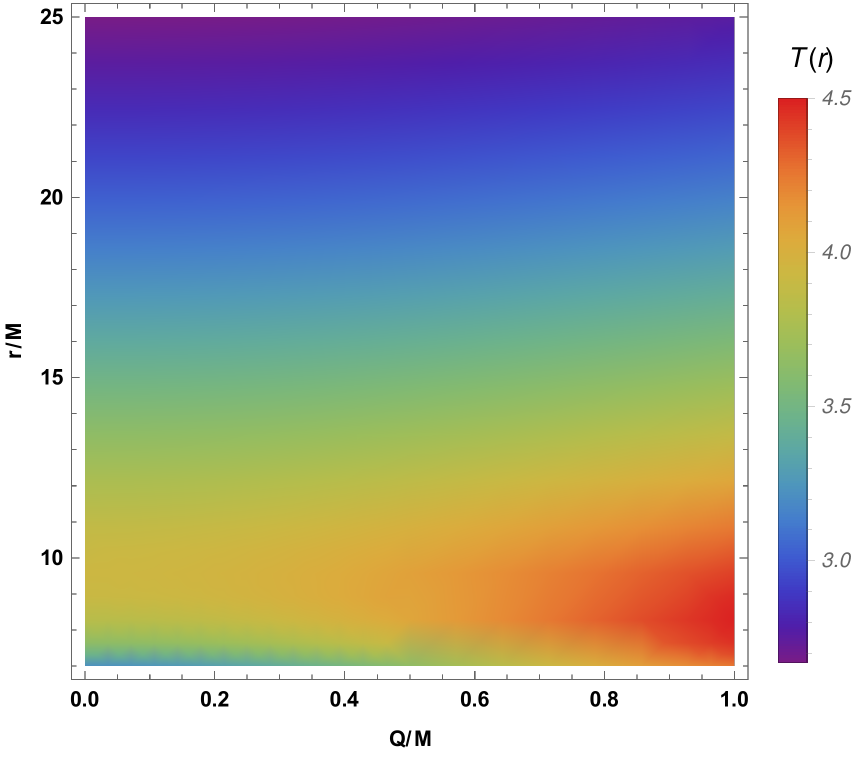}
    \includegraphics[width=0.4\textwidth]{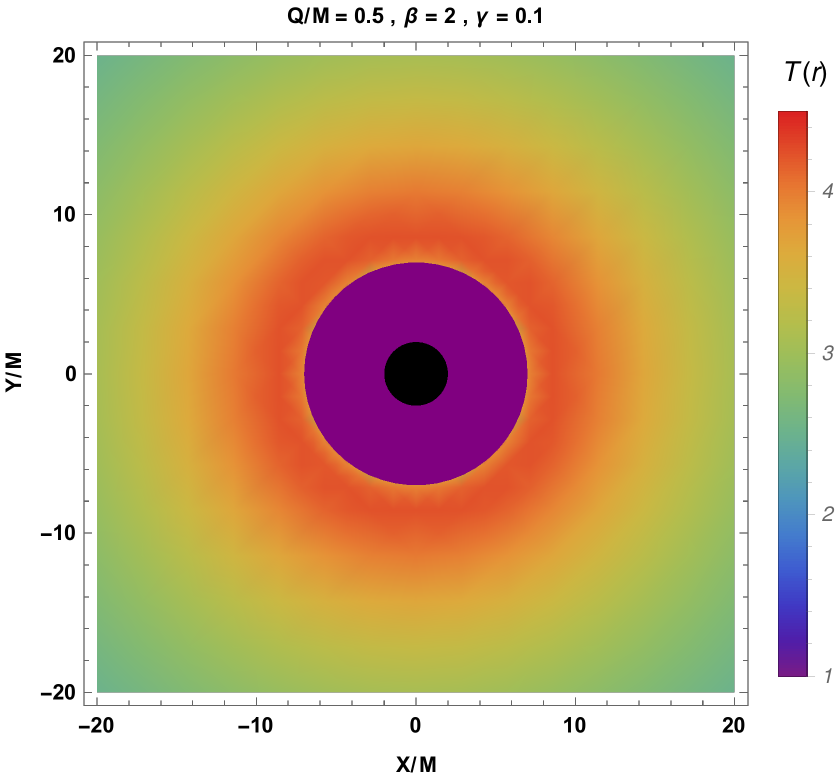}
    \caption{Left panel: temperature profile of the radial dependence of black hole charge. Right panel: temperature profile on the equatorial $X-Y$ plane in the form of density plot. Here, the Cartesian coordinates are defined by $X$ and $Y$. }
    \label{temperature2}
\end{figure*}

\section{Novikov-Thorne Model of the Accretion Disk}\label{section3}

Following the Novikov-Thorne model \cite{Novikov73}, one can let the accretion disk be optically thick and geometrically thin in the close vicinity of the EMS black hole. However, the accretion disk is really very thin in vertical size because it is extended to the horizontal size. This in turn gives rise to a negligible size in terms of the vertical size in comparison with its long horizontal extension. Therefore, its height $h$ is much smaller than the radius of the disk $r$ extended in the horizontal direction, that is,
$h \ll r$. From this property of the thin accretion disk formed in the surrounding environment of the EMS black hole and from the hydrodynamic equilibrium of the thin disk, the vertical entropy and pressure gradients are regarded as negligible in the accreting matter. The heat generated due to the stress and dynamical friction can not be accumulated in the accretion disk that is efficiently cooled as stated by the thermal radiation on the disk surface. Hence, this leads to the stabilized thin disk, whose inner edge is then located at a stable orbit around the black hole, thus taking the Keplerian motion for the plasma accreting in stable orbits.

We now define the bolometric luminosity of the accretion disk that is given by \cite{Bokhari20,Rayimbaev-Shaymatov21a}
\begin{eqnarray}
\mathcal{L}_{bol}=\eta \dot{M}c^2\, ,
\end{eqnarray}
where $\dot{M}$ and $\eta$ refer to the rate of accretion matter that falls into the black hole and the energy efficiency of the accretion disk, respectively. It is known from an astrophysical point of view that some issues exist in observing the bolometric luminosity, depending on the black hole parameters and its type. It does, therefore, become very important to measure the bolometric luminosity through the theoretical analysis and models. To this end, it is valuable to define the energy efficiency of the accretion disk around black holes, which is the highest energy that can be mined out via the the accretion disk, as that of matter falling into the black hole. Hence, 
the energy efficiency is increasingly important to explain the nature of the accretion process due to the fact that the rest mass-accreting matter turns into the electromagnetic radiation that goes out from the central object. One can then further determine the energy efficiency that stems from the radiation rate of the photon energy that is emitted from the disk surface~\cite{Novikov1973,1974ApJ...191..499P}. Thus, the efficiency can be determined via the measured energy at the ISCO \cite{sym14091765} in the case when the
photons emitted from the disk escape to infinity, that is \cite{Bardeen73} 
\begin{equation}
    \eta=1-\mathcal{E}_{ISCO}\ .
\end{equation}
The above expression for the measured energy at the ISCO allows one to find the radiative efficiency $\eta$ for the emitted photon from the disk. Therefore, it is essential to determine the ISCO parameters, that is, $r_{ISCO}$, $\mathcal{E}_{ISCO}$, and $\mathcal{L}_{ISCO}$. For that purpose, one needs to consider the test particle dynamics around electrically charged EMS black holes. It is well-known that the standard form of the effective potential for the radial motion of the test particle can generally be defined by \cite{Misner73} 
\begin{equation}
    V_{eff}(r)=-1+\dfrac{\mathcal{E}^2 g_{\phi \phi}+2 \mathcal{E} \mathcal{L} g_{t \phi}+\mathcal{L}^2 g_{tt}}{g^2_{t \phi}-g_{tt}g_{\phi \phi}}\, ,
\end{equation}
with the specific energy $\mathcal{E}=E/m$ and angular momentum $\mathcal{L}=L/m$. Here, we note that the metric functions are given by $g_{tt}=U(r)$ and $g_{\phi \phi}=f(r)\sin^2\theta$ with $g_{t\,\phi}=0$ for a static and spherically symmetric black hole. These two conserved quantities can be written in terms of metric tensors as follows:
\begin{equation}\label{energy}
    \mathcal{E}=-\dfrac{g_{tt}}{\sqrt{-g_{tt}-\Omega^2 g_{\phi \phi}}}\, , 
\end{equation}
and
\begin{equation}\label{angular}
   \mathcal{L}=\dfrac{\Omega g_{\phi \phi}}{\sqrt{-g_{tt}-\Omega^2 g_{\phi \phi}}}\, . 
\end{equation}
Similarly, the orbital angular velocity of the test particle is also given by \cite{Shapiro83,Shaymatov22a,Shaymatov22c}
\begin{equation}
    \Omega=\dfrac{d \phi}{d t}=\sqrt{-\dfrac{g_{tt,r}}{g_{\phi \phi,r}}}\ .
\end{equation}
Here we note that we further restrict motion to the equatorial plane, that is, $\theta=\pi/2$. According to the model, the accretion disk's inner edge around the black hole
can be determined by the ISCO radius that is defined by the following set of conditions for the effective potential:  
\begin{equation}
    V_{eff}(r)=0, \hspace{0.5cm} V'_{eff}(r)=0, \hspace{0.5cm} V''_{eff}(r)=0
\end{equation}

Let us then turn to the charged particle dynamics around EMS black holes. To that end, we can now define the effective potential for the charged particles moving around the EMS black hole~\cite{Kurbonov2023},
\begin{eqnarray}
V^{\pm}_{\rm eff}(r)=q A_t \pm \sqrt{U(r)\left(1+\frac{{\cal L}^2}{f(r)\sin^2\theta}\right)}\, .
\end{eqnarray}
The above effective potential has two parts, the column and gravitational interaction parts. Two different solutions exist because the effective potential can behave as symmetry in the following replacements $qQ \to -qQ$: $V^+_{\rm eff} \to -V^-_{\rm eff}$ and $V^-_{\rm eff} \to -V^+_{\rm eff}$. 
However, we will restrict ourselves to the
effective potential $V^+_{\rm eff}$ that corresponds to the positive energy for physically meaningful timelike motion and for analysing the charged particle's dynamics. Taking $V^+_{\rm eff}$ into consideration, we can further find the ISCO values of the charged particle, that is, $r_{ISCO}$, $\mathcal{E}_{ISCO}$, and $\mathcal{L}_{ISCO}$. It is worth noting that we decide to explore the ISCO values numerically. The results are presented in Table~\ref{Tab2}. As can be observed from the results on the ISCO radius determining the inner edge of the accretion disk, it decreases as a consequence of an increase in the value of the dimensionless parameter $\beta$, whereas it increases under the effect of parameter $\gamma$. Taking all together, we then analyse the energy efficiency of the accretion disk around the black hole in EMS theory. The value of the energy efficiency is presented in Table \ref{Tab2}. As can be observed from the results, the radiative efficiency increases with an increase in the value of parameter $\beta$, whereas it decreases as a consequence of the impact of parameter $\gamma$. Note that the Schwarzschild black hole case can be obtained in the limit of $\beta=0$ and $\gamma=0$. At this point, the efficiency is equal to $\eta \approx 5.7 \%$ \cite{Kurmanov_2022}. Similarly, the following limit, $\beta=1$ and $\gamma=0$, refers to the RN black hole case. The radiative efficiency is then evaluated to be $\eta \approx 7.5 \%$ in the RN black hole case.
\begin{table*}[ht]
\begin{center}
\caption{Value of the innermost stable circular orbit (ISCO) radius and radiative efficiency of the accretion disk tabulated for the different values of parameter $\beta$ in the possible cases of parameter $\gamma$. Note that we have set $q=0.5$ and $Q=0.5$.}\label{Tab2}
\resizebox{.7\textwidth}{!}
{
\begin{tabular}{l l|l l |l l| l l  }
 \hline \hline
 & &   $\gamma=-1$  & &$\gamma=0$  & &$\gamma=1$    
 \\
\cline{3-5}\cline{6-8}
& $\beta$    & $r_{ISCO}$ & $\eta \%$   & $r_{ISCO}$ & $\eta \%$ & $r_{ISCO}$ & $\eta \%$   \\
\hline
& 0.0   & 5.76641 & 7.50318  & 6.0  & 5.7191 & 6.2865  &4.0603  \\
& 0.5   & 5.61325 & 8.41598  & 5.82216  & 6.58717 & 6.05386  &4.89047 \\
& 1.0     & 5.46477 & 9.35293 & 5.65896  & 7.47554 & 5.85509 & 5.74132  \\
& 2.0       & 5.16738 & 11.3155 & 5.35546  & 9.31973 & 5.51641 & 7.50318  \\
 \hline \hline
\end{tabular}
}
\end{center}
\end{table*}

\section{Radiative Properties of the Accretion Disk around a Central Compact Object}\label{section4}

In this section, we study the flux produced by an accretion disk of the EMS black hole. We note that an accretion disk consists of gas and dust moving on stable orbits of a black hole or neutron star. The gas and dust can lose their energy and angular momentum owing to the background geometry when they start moving around the compact object. As a result, the orbits of gas and dust move to the inner edge of the accretion disc so that they fall towards the compact object. This process ends up with accretion disk radiation, in which the gas and dust can heat up and emit radiation. We further expect that the accretion disk radiation can be affected by the EMS black hole parameters. In particular, the black hole charge can affect the gas and dust in the accretion disk around the black hole, thus resulting in highly ionized gas and dust in the accretion disk around the black hole. As a consequence of the ionization process, the high energy radiation can be emitted in the form of X-rays that can be detected. 
The radiation of the accretion disk around a black hole in the EMS theory is expected to be a potent test to explain valuable and remarkable aspects of the accretion disk and can also lead to serious implications from the observational point of view. The flux of the electromagnetic radiation can be defined by the following equation~~\cite{Novikov1973,Shakura:1972te,Thorne:1974ve}:
\begin{equation}\label{flux}
    \mathcal{F}(r)=-\dfrac{\dot{M_0}}{4 \pi \sqrt{g}}\dfrac{\Omega_{,r}}{(E-\Omega L)^2} \int _{r_{ISCO}}^r (E-\Omega L) L_{,r} d r\, , 
\end{equation}
with the determinant of the three-dimensional subspace $g=\sqrt{-g_{tt}g_{rr}g_{\phi \phi}}$ with ($t, r, \phi$) coordinates. In the above equation, the quantity $\mathcal{F}(r)$ can be defined as the function of $\dot{M_0}$ that can be regarded as unknown and corresponds to the accretion rate of the disk mass. However, we can set $\dot{M_0}=1$ for further analysis. 

Using Eqs.~(\ref{energy}) and (\ref{angular}) we further derive the flux of the electromagnetic radiation. It does however turn out that it is complicated to derive the analytical expression of the flux, and thus we resort to it numerically; see Fig.~\ref{fig:flux}. As can be observed from Fig.~\ref{fig:flux}, the shape of the flux of the electromagnetic radiation increases/decreases under the effect of the $\gamma$ parameter as compared with the RN case, depending on its sign. However, it increases as a consequence of an increase in the value of black hole charge $Q$ and parameter $\beta$. In the top-left panel of Fig.~\ref{fig:flux}, we show the flux of the electromagnetic radiation for the Schwarzschild and RN black hole cases. From Fig.~\ref{fig:flux}, it can be easily observed that the black hole charge increases the value of the flux of the electromagnetic radiation. In addition, the flux of the black body radiation can be written as $\mathcal{F}(r)=\sigma T^4$ with $\sigma$ as the Stefan-Boltzmann constant. The radial dependence of the disk temperature and energy is further shown in Fig.~\ref{temperature1} for various values of black hole charge. In this graph, we select $\beta=0$ \& $\gamma=0$ and $\beta=1$ \& $\gamma=0$ for the Schwarzschild and Reissner-Nordstr\"{o}m black holes, respectively. As can be observed from Fig.~\ref{temperature1}, the temperature and energy of the accretion disk around the EMS black hole increase with increasing black hole charge $Q$.
\begin{figure}[ht]
    \centering
    \includegraphics[scale=0.5]{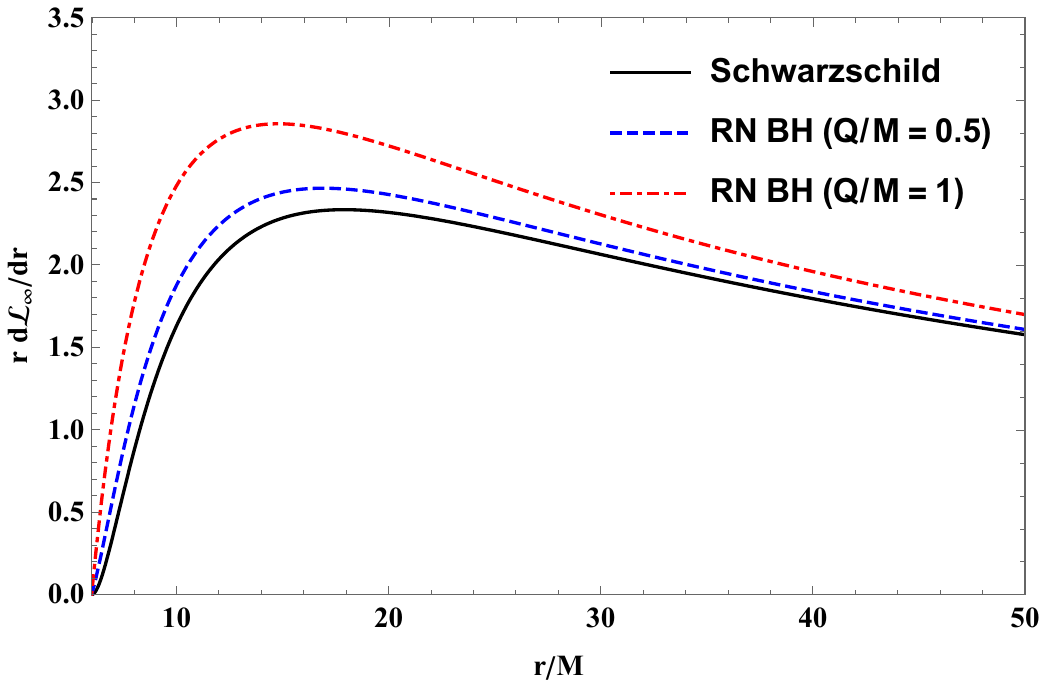}
    \caption{Radial dependence of the differential luminosity of the accretion disk scaled in powers of $10^{-2}$ plotted as a result of the numerical evaluation.}
    \label{fig:luminosity}
\end{figure}

To be more informative, we plot the temperature through by "color map" in Fig.~\ref{temperature2}. From Fig.~\ref{temperature2}, it can be clearly observed that the dark region lies inside the accretion disk's inner edge, whereas the red parts refer to the regions where the temperature of the disk can reach its maximum value. 

We then turn to another important quantity referred to as the differential luminosity. It can be expressed as follows:~\cite{Novikov1973,Shakura:1972te,Thorne:1974ve}
\begin{equation}\label{Eq:luminosity}
    \dfrac{d \mathcal{L}_{\infty}}{d \ln{r}}=4 \pi r \sqrt{g} E \mathcal{F}(r)\, .
\end{equation}
We assume that the radiation emission can be described by black body radiation. Keeping this in mind, one can then define the spectral luminosity $\mathcal{L}_{\nu,\infty}$ as a function of the radiation frequency $\nu$ at infinity as follows:~\cite{Boshkayev:2020kle,sym14091765,Shaymatov2023}
\begin{equation}\label{luminosity2}
    \nu \mathcal{L}_{\nu,\infty}=\dfrac{60}{\pi^3} \int_{r_{ISCO}}^{\infty} \dfrac{\sqrt{g} E}{M_T^2}\dfrac{(u^t y)^4}{\exp\Big[{\dfrac{u^t y}{(M_T^2 \mathcal{F})^{1/4}}}\Big]-1}\, ,
\end{equation}
with $y=h \nu /k T_{\star}$, where $h$ and $k$ respectively refer to the Planck constant and constant of Boltzmann, as well as the total mass $M_T$. In addition, $T_{\star}$ corresponds to the characteristic temperature and has a relation with the Stefan-Boltzmann law, that is, $\sigma T_{\star}= \dfrac{\dot{M}_0}{4 \pi M_T^2}$, where $\sigma$ accordingly refers to the Stefan-Boltzmann constant. Another important quantity is the differential luminosity. The radial dependence of the differential luminosity is represented in Fig.~\ref{fig:luminosity}. The electromagnetic radiation flux of the accretion disk can be scaled in powers of $10^{-5}$. The differential luminosity of the accretion disk depends on the flux by the relation given in Eq.~(\ref{Eq:luminosity}). Therefore the value of the differential luminosity can be of the order of $10^{-2}$.
As can be observed, the value of the differential luminosity increases as a consequence of an increase in the value of black hole charge $Q$.

\section{Conclusions}\label{summary}
In this paper, we studied the radiation properties of the accretion disk around black holes in EMS theory as the accretion disk radiation can be affected by the EMS black hole parameters. The theoretical study of the accretion disk radiation around a black hole in EMS theory is expected to be a potent test to explain its valuable properties and the astrophysical observations. We can summarize the main results of our research as follows:
\begin{itemize}
    \item 
    We extensively studied the behavior of metric functions in Fig.~\ref{fig:metric} and we found the horizon behaviour of the black hole in EMS theory. In addition, 
    we obtained the extreme values of the black hole parameters using the equation for the horizon. We present the results for the extreme value of black hole parameters in Table~\ref{Tab1}.

    \item
    We also investigated the effective potential for the charged particles. On the basis of the effective potential, we further found the ISCO radius numerically and present the results in Table \ref{Tab2}. We found that parameters $\gamma$ and $\beta$ have significant influence on the ISCO radius of the charged particles.

    \item 
    Further,  we explored the radiative efficiency and present the results in Table \ref{Tab2}. It was shown that the radiative efficiency increases with increasing parameter $\beta$, while it decreases as a consequence of an increase in the value of parameter $\gamma$. 

    \item 
    Finally,  we considered the radiative properties of the black hole accretion disk, that is, we explored the flux, temperature, and spectral and differential luminosities of the black hole accretion disk. We show the results in Figs.~\ref{fig:flux}-\ref{fig:luminosity}. It was found that the black hole electric charge increases the flux of the electromagnetic radiation in EMS theory. Similarly, the temperature and energy of the accretion disk increase due to the impact of the black hole charge, as can be observed in the graph of temperature in the form of a density-plot in Fig.~\ref{temperature2}. 
\end{itemize}

\section{Acknowledgments}
This work is supported by the National Natural Science Foundation of China under Grants No. 11675143 and No. 11975203, the National Key Research and Development Program of China under Grant No. 2020YFC2201503. M.A. and B.A. wish to acknowledge the support from Research F-FA-2021-432 of the Uzbekistan Agency for Innovative Development.

\bibliography{ref}

\end{document}